\documentclass[useAMS,usenatbib]{mn2e}

\newcommand{\msun}{M$_{\sun}$}

\newcommand{\msuns}{M$_{\sun}~$}

\newcommand{\mnras}{MNRAS}
\newcommand{\apj}{ApJ}
\newcommand{\apjl}{ApJ}
\newcommand{\aj}{AJ}
\newcommand{\aap}{A\&A}
\newcommand{\araa}{ARA\&A}
\newcommand{\apss}{APSS}

\usepackage{graphicx}

\title[Does sub-cluster merging accelerate mass segregation in local star formation?]{Does sub-cluster merging accelerate mass segregation in local clusters?}
\author[N. Moeckel and I. A. Bonnell]{Nickolas Moeckel$^{1}$\thanks{E-mail:
nickolas1@gmail.com} and Ian A. Bonnell$^{2}$\\
$^{1}$School of Physics, University of Exeter, Stocker Road, Exeter, EX4 4QL\\
$^{2}$SUPA, School of Physics and Astronomy, University of St Andrews, North Haugh, St Andrews, Fife, KY16 9SS
}
\begin{document}

\date{Submitted 25 June, 2009}

\pagerange{\pageref{firstpage}--\pageref{lastpage}} \pubyear{2009}

\maketitle

\label{firstpage}

\begin{abstract}
The nearest site of massive star formation in Orion is dominated by the Trapezium subsystem, with its four OB stars and numerous companions.  The question of how these stars came to be in such close proximity has implications for our understanding of massive star formation and early cluster evolution.  A promising route toward rapid mass segregation was proposed by \citet{mcmillan07}, who showed that the merger product of faster-evolving sub clusters can inherit their apparent dynamical age from their progenitors.  In this paper we briefly consider this process at a size and time scale more suited for local and perhaps more typical star formation, with stellar numbers from the hundreds to thousands.  We find that for reasonable ages and cluster sizes, the merger of subclusters can indeed lead to compact configurations of the most massive stars, a signal seen both in Nature and in large-scale hydrodynamic simulations of star formation from collapsing molecular clouds, and that sub-virial initial conditions can make an un-merged cluster display a similar type of mass segregation.   Additionally, we discuss a variation of the minimum spanning tree mass-segregation technique introduced by \citet{allison09a}.  
\end{abstract}


\section{Introduction}
The question of how dense, massive groupings like the Trapezium are formed has yet to find a convincing answer.  Ignoring for the moment the striking multiplicity of the system\footnote{Though a complete theory for the formation of a Trapezium-like system needs to include this key aspect.} \citep{preibisch99,schertl03,kraus07}, a first-order glance at the Trapezium shows the most massive stars in the cluster arranged in a central, compact configuration.    Observations of a possible proto-trapezium, W3 IRS 5 \citep{megeath05}, seem to show a massive subcluster still in the embedded phase, which could be suggestive of either formation as a compact cluster or gas-driven migration during formation.  However, if a compact system of massive stars were to form {\it in situ}, it would be dynamically unstable \citep{pflamm-altenburg06}; if the stars in the Trapezium formed in such a fashion, perhaps an order of magnitude more stars were originally in the system in order to leave behind the 4 OB stars today.  Alternatively, the stars may have migrated there by some combination of dynamical mass segregation or gas dynamical effects during formation.  The problem with dynamical segregation alone (as opposed to gas-driven segregation) is the young age of the cluster and the segregation timescale, which taken together at face value would suggest that the cluster is not old enough to have segregated to the degree seen \citep{bonnell98}. 

Recent work \citep{mcmillan07} has suggested a scenario in which merging smaller-{\it n} subclusters can create a situation in which the final larger-{\it n} cluster appears to be dynamically much older than the age of the stars would suggest.  The key concept is that dynamical mass segregation occurs on a shorter timescale in the low-{\it n} subclusters, and that the segregation is maintained in the merger process, so that effects of the shorter dynamical timescale of the subclusters are applied to the structure of the final merger products.  This idea is supported by phase-space arguments by \citet{fellhauer09}.  This sort of later dynamical mixing is perhaps supported by observations suggesting that young massive stars are frequently found in close proximity to other stars of varying ages \citep{brogan08}.

The McMillan et al. results point the direction toward an attractive solution to the rapid concentration of massive stars as seen in the Trapezium and W3 IRS 5.  However, their simulations employed large numbers of stars, with $n = 10^4$--$4\times10^4$.  It remains untested\footnote{Although after submission of this paper, \citet{allison09b} published a similar study to this one, which we address in the discussion.} how applicable these results are to less prolific but more representative star clusters with $n=300$--1000, which may be more representative of a `typical' star forming region \citep{lada03,porras03}.  In these smaller clusters with grainier potentials and crossing times that are smaller than their larger siblings, it is unclear if the factor of $\sim 2$ decrease in the dynamical time gained by halving the number of stars will come into play.  In this paper we investigate the merger of low-{\it n} sub-clusters to determine if this process is likely to play a role in accelerating the concentration of massive stars in young local clusters.

\section{Measuring segregation}
There are several ways to measure the level of segregation in a cluster, including the evolution of the Lagrangian radii, the half-mass radii of different mass components, or their cumulative number distributions.  A new method was recently developed by \citet{allison09a}, utilizing the minimum spanning tree (MST) of the stars of interest relative to that of a random subset of the cluster stars.  An attractive feature of this method is that it doesn't rely on defining the center of a cluster.  The method we use here is directly inspired by the MST method.

Rather than use the MST, we use the two-dimensional convex hull of the stars' positions projected onto the plane.  The convex hull encloses the minimal area that contains all straight lines joining any two points; in two dimensions it is conceptually found by stretching an elastic band around the points of interest and letting it relax, and it is unique for a given set of points.  The area of the convex hull of $N$ stars, which we denote $H_c(N)$, divided by $N$ yields an effective surface density for the stars, and can be used like the MST to measure the compactness of a subset of stars in a cluster.

\begin{figure}
 \includegraphics[width=84mm]{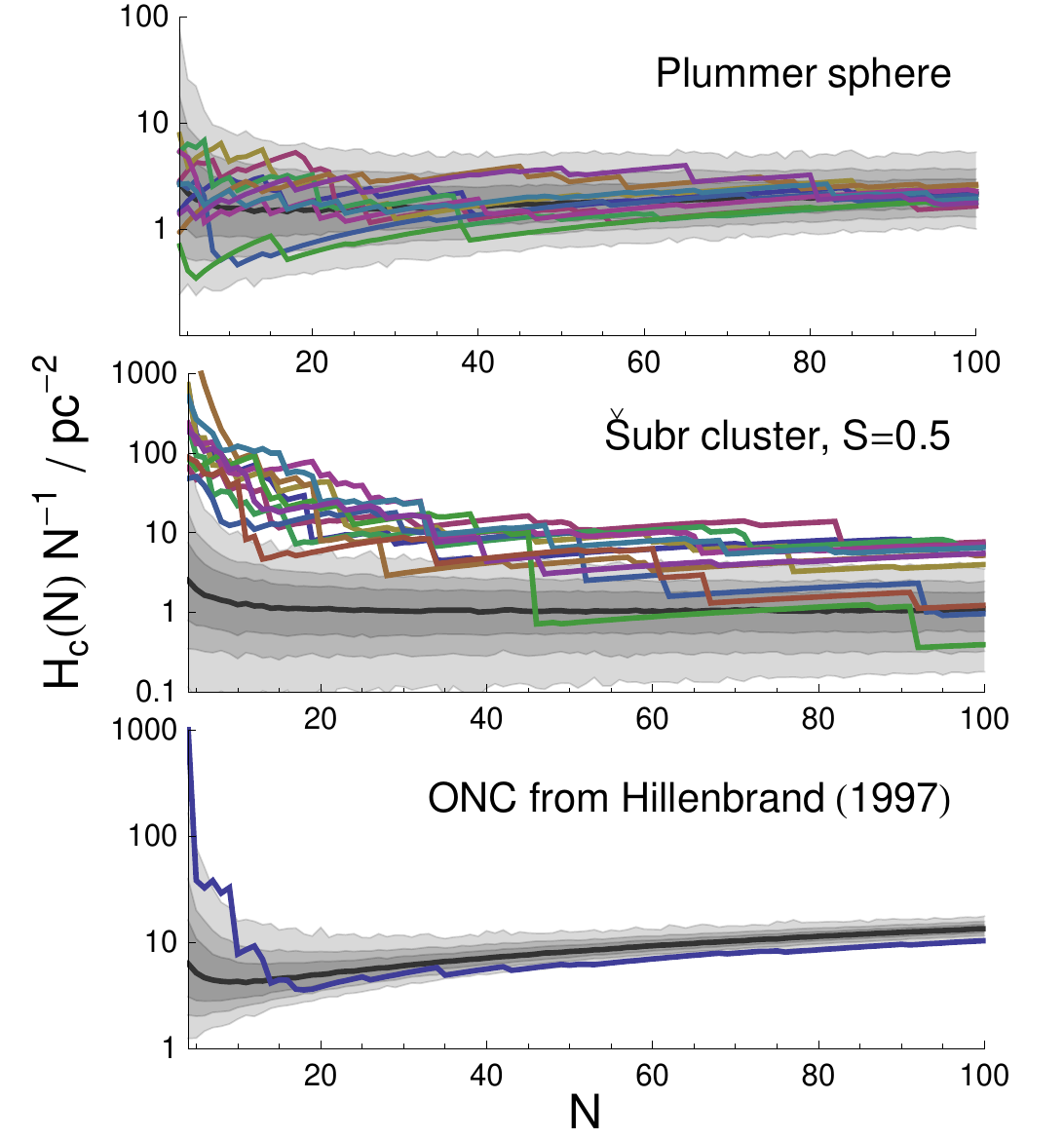}
 \caption{Examples of mass segregation as detected by the scheme used in this work.  In each plot the black line is the median value of $H_c(N)$ for the given $N$, and the progressively lighter gray regions show the $\pm1\sigma$, $\pm 2\sigma$, and $\pm 3\sigma$ regions.  The coloured lines show 10 realizations of a random setup for the Plummer and {\v S}ubr clusters, and for the ONC case the data from \citet{hillenbrand97}.  The Plummer spheres and {\v Subr} clusters all have 1000 stars and half-mass radii $\sim 0.5$ pc.  \label{ConvexHullExamples}}
\end{figure}

Our scheme proceeds much as the MST method, in the following manner for the $N$ most massive stars in the cluster: first, compute $H_c(N)$ for many subsets of $N$ stars drawn randomly from the cluster.  Computing the convex hull is not intensive\footnote{There are several algorithms for the calculation \citep[e.g.][]{press07}, which in two dimensions scales as $n{\rm log} (n)$ for $n$ points.  This work uses the built-in function implemented in {\em Mathematica}.}; $10^4$ subsets of 100 stars can be computed in $\sim 15$ seconds on an aging Apple MacBook with a 2 GHz Intel Core 2 Duo and a loud cooling fan.  After computing 1000 or more values of $H_c(N)$, the median and $\pm1\sigma$, $\pm 2\sigma$, and $\pm 3\sigma$ values of the distribution can be determined.  These define the `normal' range of values for $H_c(N)$ in the cluster for a {\em random sampling} of $N$ stars.  Second, compute $H_c(N)$ for the $N$ {\em most massive} stars.  Finally, repeat these steps for many values of $N$ up to the highest interesting value.  The compactness of the $N$ most massive stars can then be compared to the expected value for that $N$, to see if they are more closely clustered than a random placing would be likely to yield.

\begin{figure}
 \includegraphics[width=84mm]{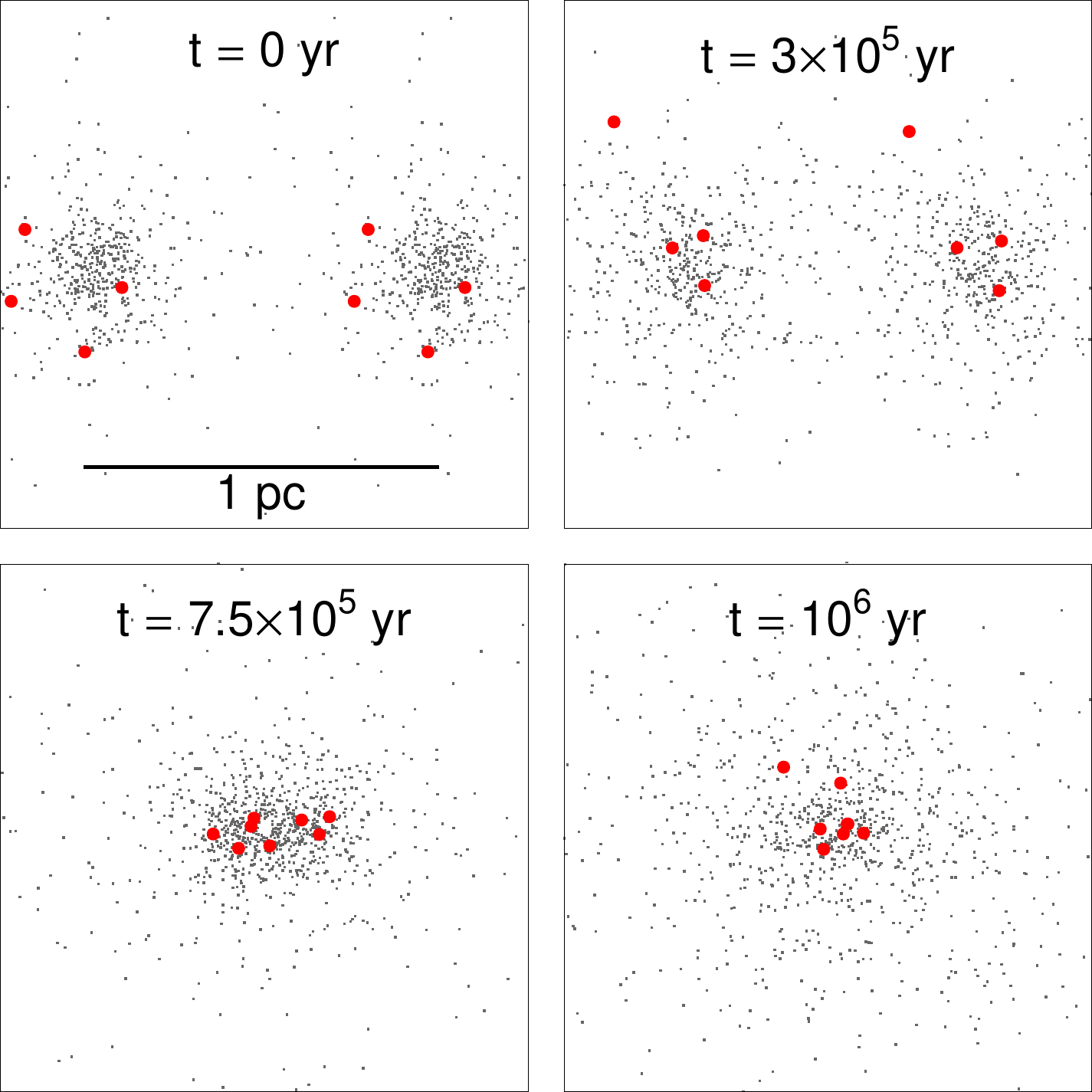}
 \caption{Example of two merging clusters, each with $n=500$.  Red dots show the positions of the eight most massive stars in the experiment.\label{MergeExample}}
\end{figure}

To illustrate this scheme, we use it to compute the segregation levels for three test scenarios: an unsegregated Plummer sphere; a segregated cluster using the method of \citet{subr08}, using their value $S=0.5$ corresponding to a dynamically mature cluster; and following \citet{allison09a}, the Orion Nebula Cluster using the data from \citet{hillenbrand97}.  The Plummer sphere and segregated clusters all have 1000 stars, and characteristic radii $\sim 0.5$ pc.  These examples are shown in figure \ref{ConvexHullExamples}.  In each panel, the black line is the median value of $H_c(N)$ for 3000 random samples of $N$ stars, and the progressively lighter gray regions show the $\pm1\sigma$, $\pm 2\sigma$, and $\pm 3\sigma$ values of this distribution.  These regions are to be compared to $H_c(N)$ of the $N$ most massive stars, which we plot as coloured lines.

Looking at the Plummer sphere in the top panel, note that most of the coloured lines, showing the values of 10 individual random realizations of the cluster setup, lie within the $\pm2\sigma$ shaded region of typical surface densities for each $N$.  This is then the indication of an unsegregated cluster.  Note that the median values and shaded regions are taken from a single realization; for these clusters and for the segregated cluster, the spatial distributions of each cluster are very similar and the median and shaded regions are almost indistinguishable between realizations.  In the middle panel we show a heavily segregated cluster, and we see that for all cases $H_c(N)$ of the $N$ most massive stars lies above the $+2\sigma$ value up to $N=45$, and for seven out of the ten this extends out past $N=100$.  This means that for most of the realizations of this setup, at least the 100 most massive stars are distributed in a fashion that is more compact than a random sampling of 100 stars would yield.

Turning to the  lower panel, we show the method as applied to the data from \citet{hillenbrand97} for the ONC.  We see that for the lowest value of $N$ that we consider, $N=4$, the surface density of the most massive stars is quite high.  For $N <10$, $H_c(N)$ lies around or above the $+3\sigma$ value of the cluster distribution, before falling back at or below the cluster median for higher values of $N$.  This agrees quite well with the result \citet{allison09a} found for the same data, which is unsurprising given that this method is fundamentally indebted to theirs.  Note also the higher values of the surface density for the ONC compared to the randomly generated clusters.  This illustrates the retention of physical meaning this method offers, showing both the absolute surface density and the relative amount of mass segregation for the most massive stars in a cluster.

\begin{figure*}
 \includegraphics[width=160mm]{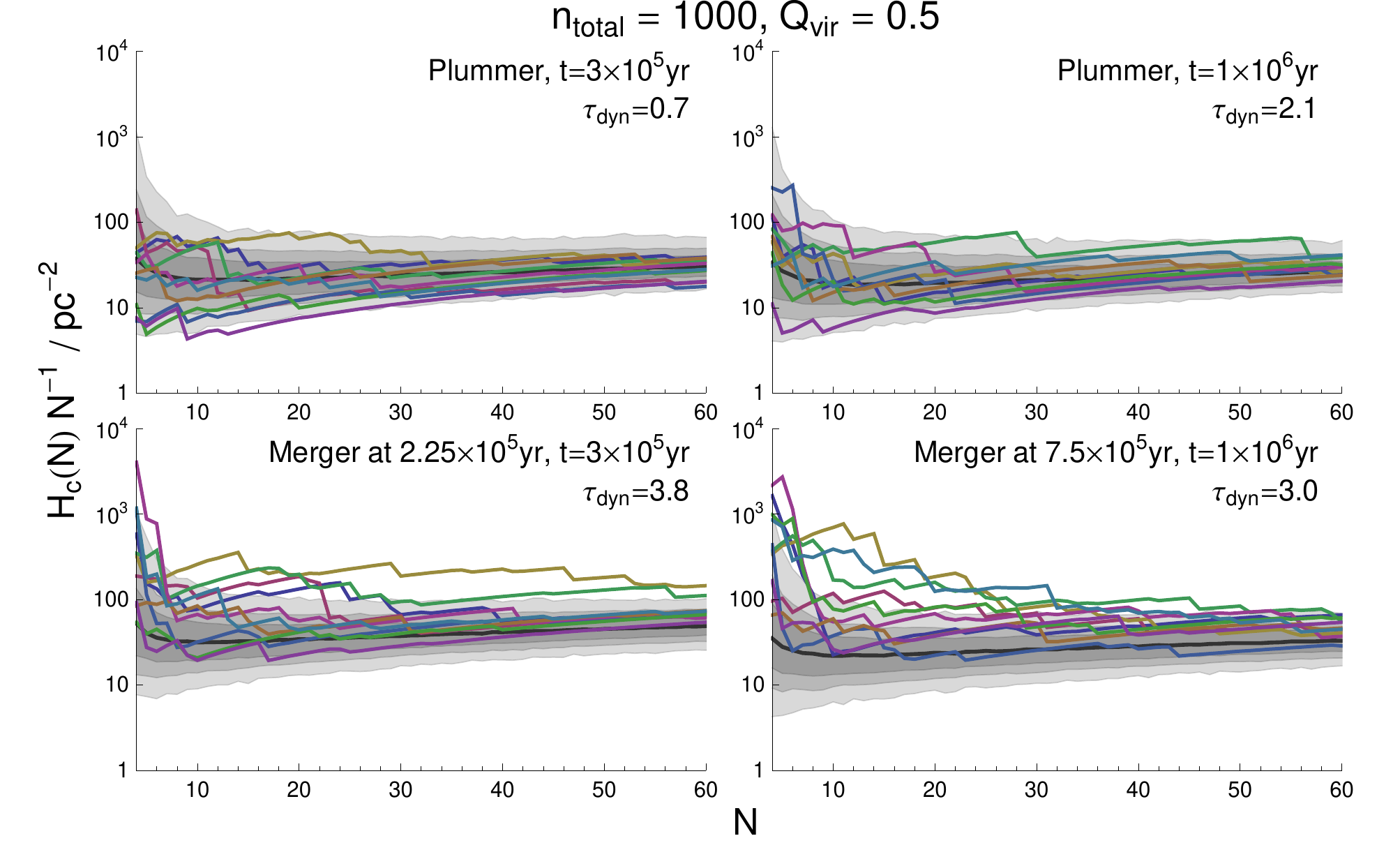}
 \caption{Mass segregation from our runs with 1000 stars total, with virialized initial conditions.  As in figure \ref{ConvexHullExamples}, the coloured lines are 10 random realizations of the initial conditions, and the gray shaded regions show the $\pm1, \pm2$, and $\pm3$ sigma values of $H_c(N)$ about the median black line.  Time in years and the apparent dynamical age are labeled.  \label{1000synthesisvirial}}
\end{figure*}

\begin{figure*}
 \includegraphics[width=160mm]{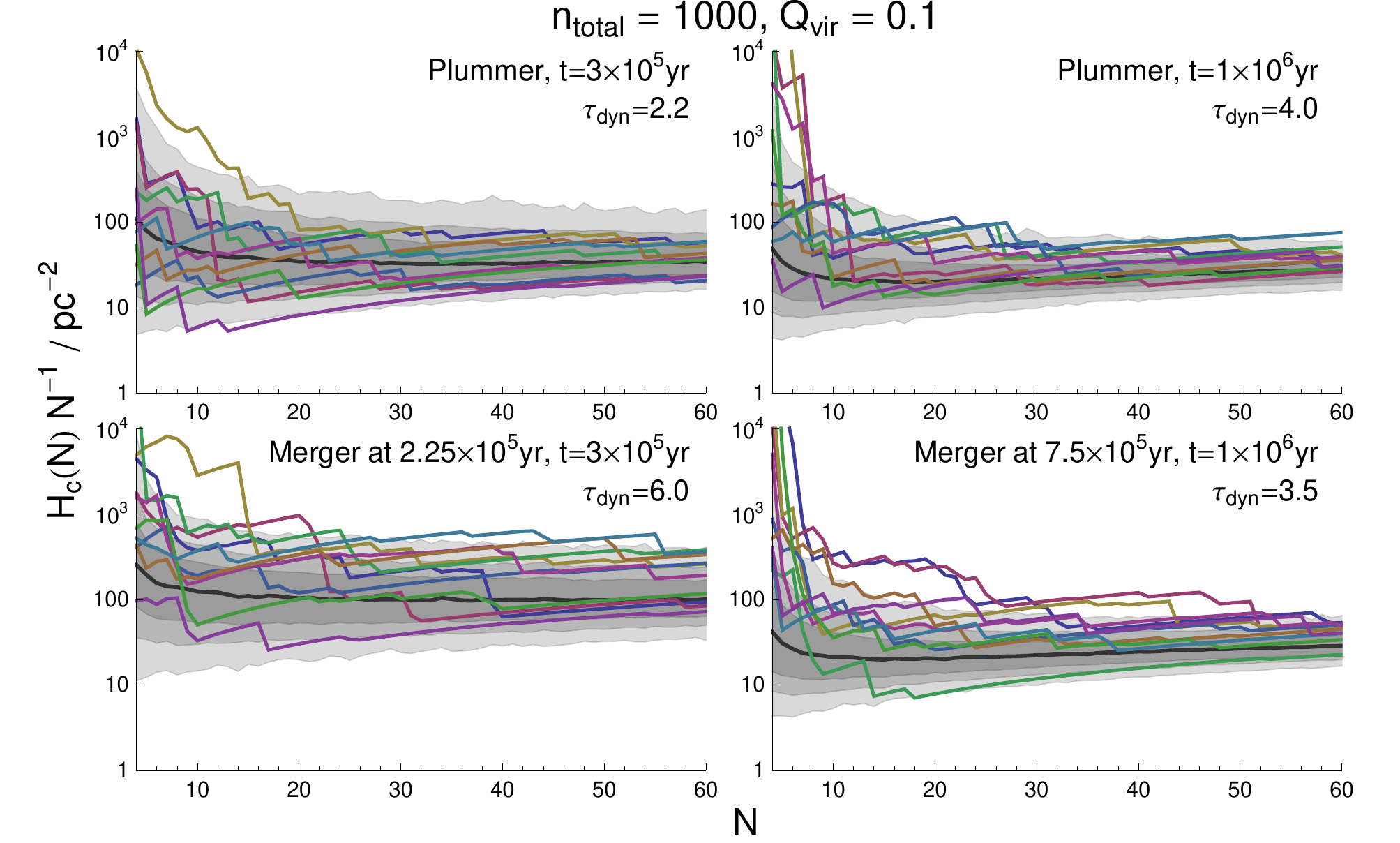}
 \caption{As figure \ref{1000synthesisvirial}, for subvirial initial velocities.  \label{1000synthesiscold}}
\end{figure*}

\begin{figure*}
 \includegraphics[width=160mm]{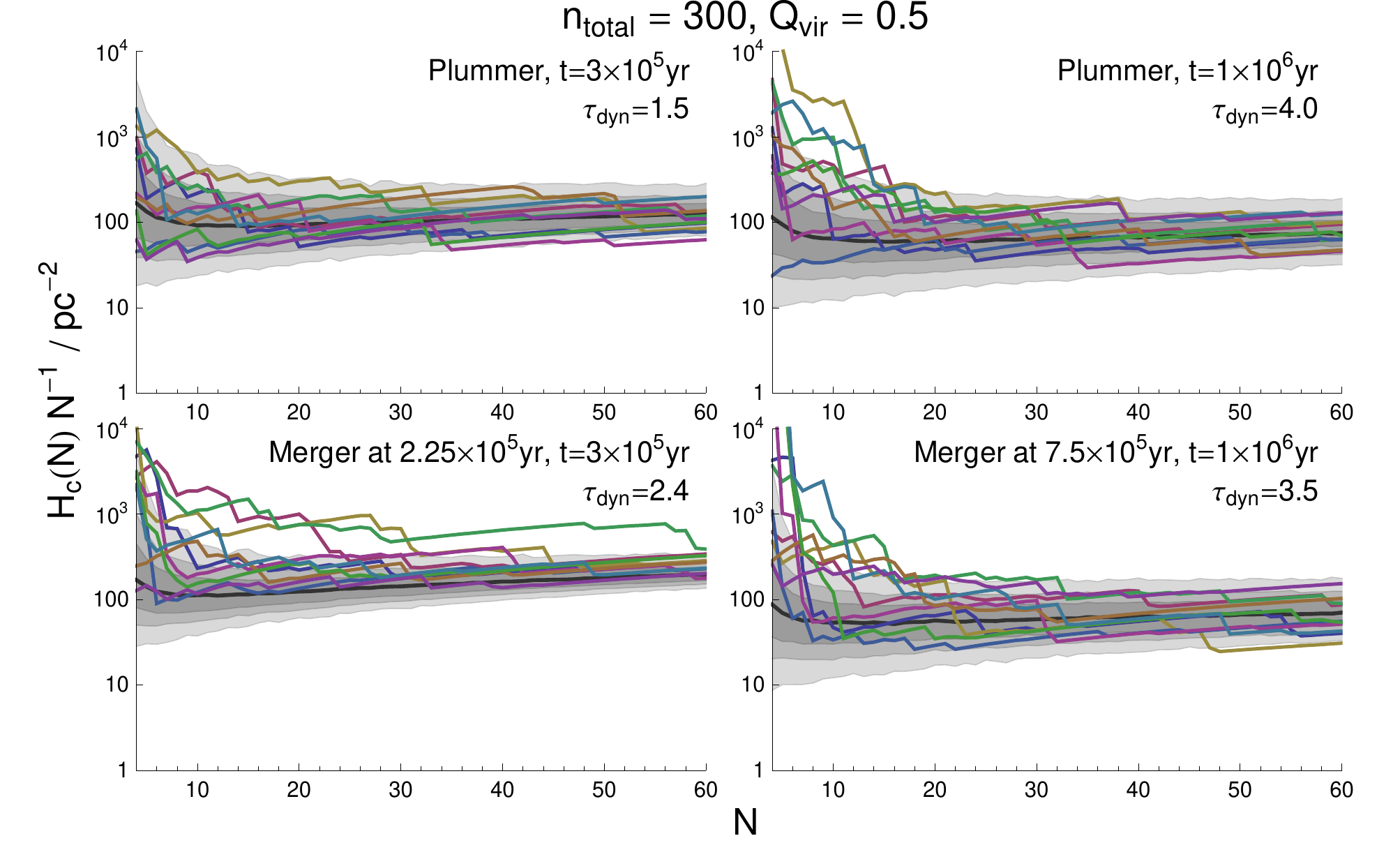}
 \caption{Mass segregation from our runs with 300 stars total, with virialized initial conditions.  As in figure \ref{ConvexHullExamples}, the coloured lines are 10 random realizations of the initial conditions, and the gray shaded regions show the $\pm1, \pm2$, and $\pm3$ sigma values of $H_c(N)$ about the median black line.  Time in years and the apparent dynamical age are labeled.  \label{300synthesisvirial}}
\end{figure*}

\begin{figure*}
 \includegraphics[width=160mm]{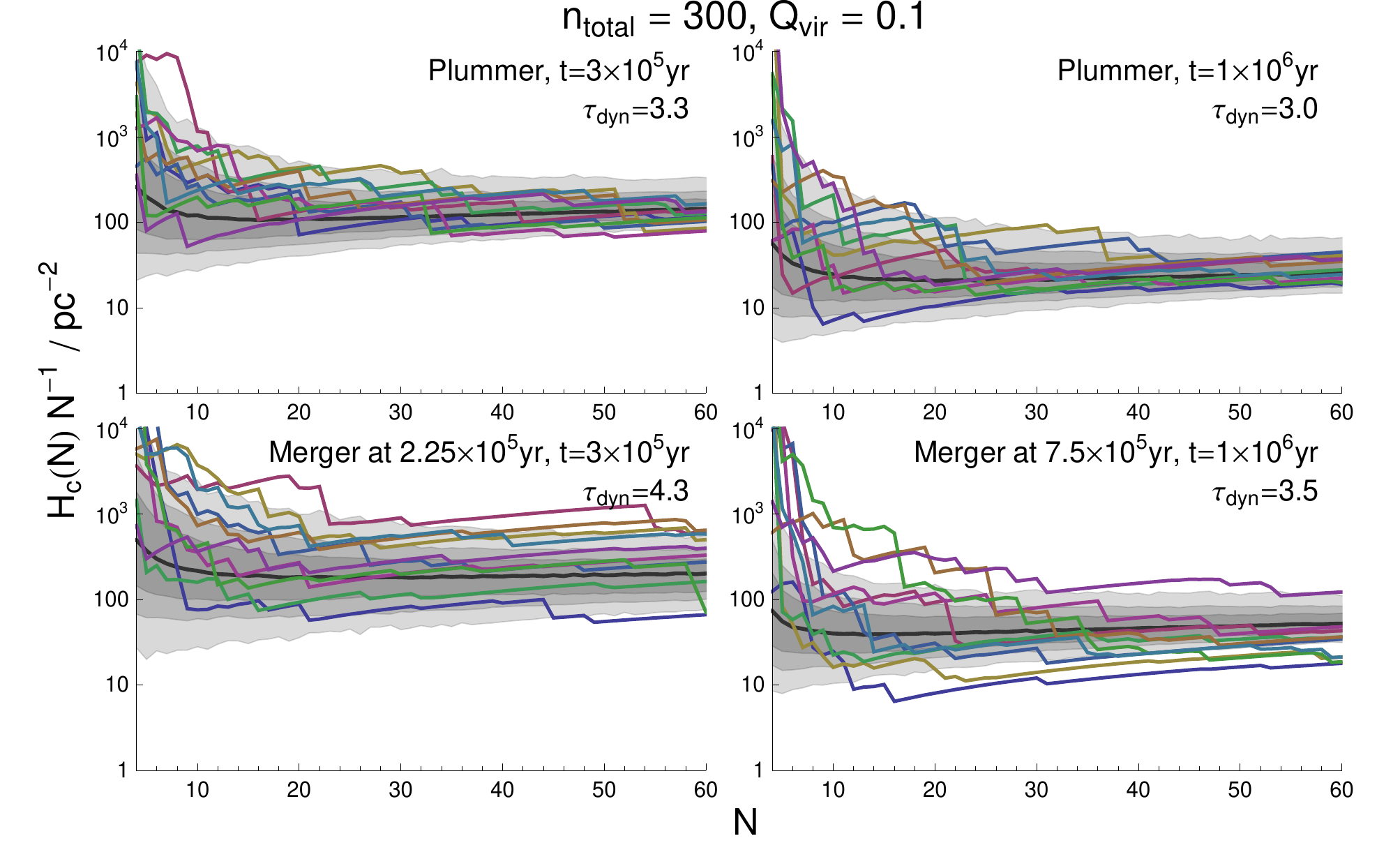}
 \caption{As figure \ref{300synthesisvirial}, for subvirial initial velocities.  \label{300synthesiscold}}
\end{figure*}
\section{Numerical experiments}

We performed experiments to compare the amount of mass segregation obtained with a single Plummer sphere to that resulting from the merger of two smaller Plummer spheres.  The systems were integrated using {\sc nbody6} \citep{aarseth03}.  As these are merely illustrative experiments, we used a simple one-component mass function with masses in the range 0.1 -- 10.0 \msun and a power-law index $\alpha = -2.3$.  We ran cases with two final cluster sizes: larger clusters where the single cluster had $n = 1000$ and scale length $a=0.5$ pc, and the merging clusters had $n = 500$ and $a=0.25$ pc; and smaller clusters where the single sphere had $n=300$ and $a=0.2$ pc, and the merging clusters had $n=150$ and $a=0.1$ pc.  These cluster sizes are a factor of a  $\sim$2--4 times more compact then the clusters in \citet{lada03}\citep[see figure 2 of][]{adams06}.  If one considers the two merging spheres as a single cluster at a pre-merger time, however, they would be larger than that sample.  This is also ignoring the (unmodeled here) effects of gas dispersal and non-sphericity.  We chose the upper mass limit of 10 \msuns as we are interested primarily in low-to-moderate mass star formation.  Additionally, our clusters are of relatively low mass ($\sim 100$--325 \msun), where one would arguably not expect to find a star of much greater mass than 10 \msuns \citep{weidner06}.

The merging clusters were created by setting up one Plummer sphere and setting its virial ratio, then duplicating the cluster.  The copies were initially at rest with respect to each other, placed so that they merged at an `early' time, $2.25\times10^5$ yr, and a `late' time, $7.5\times10^5$ yr.  Segregation was then measured at $3\times10^5$ yr and $10^6$ yr for both the merger product and the single Plummer sphere.  An example of this is shown in figure \ref{MergeExample}.  As there is some evidence that star formation may result in sub-virial stellar velocities \citep[e.g.][]{walsh04,peretto06,kirk07}, each setup was run from virialized and cold initial conditions, with virial parameters $Q_{vir} = 0.5$ and 0.1.  Each set of parameters was run with 10 random realizations of the initial conditions.  As a simple way to exclude escaping stars from the mass segregation calculation, we restrict our analysis of the results to stars that remain within 1.5 pc of the center of mass.

\begin{figure*}
 \includegraphics[width=160mm]{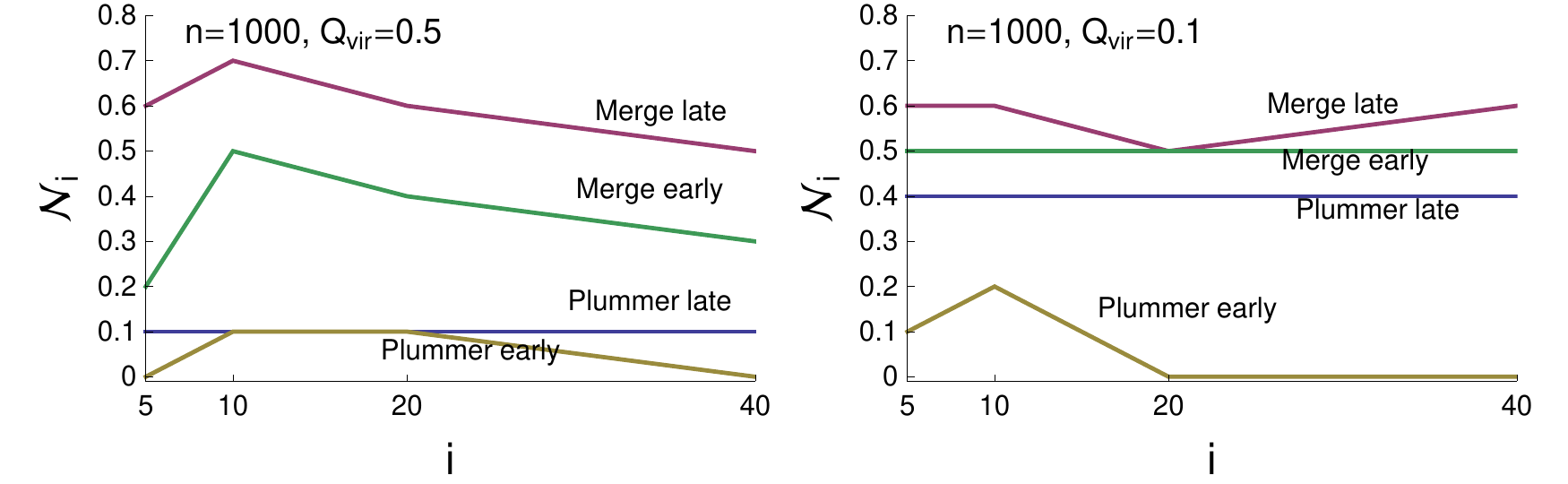}
 \caption{The fraction of runs ${\mathcal N}_i$ showing segregation for the $i$ most massive stars, for the $n=1000$ clusters.\label{tableexample}}
\end{figure*}

We have anchored our experiments in physical time from the commencement of the integrations, rather than in terms of the initial crossing time, which only has observational significance to the single Plummer spheres with their relatively static structure.  We therefore also determine for each set of parameters an {\em apparent} dynamical age, which is what a present-day observer might measure (assuming reasonably accurate stellar ages can be determined).  We calculate this by determining the crossing time at the {\em end} of the experiment and dividing this by the age of the system in years, i.e. $\tau_{dyn} = t_{yr}/t_{cross, final}$.  We define the crossing time as $t_{cross} = 2 r_h / v_{rms}$, with $r_h$ the half-mass radius and $v_{rms}$ the root mean squared velocity.  We also show the age of the system in terms of its final relaxation time, $\tau_{relax} = t_{yr}/t_{relax,final}$, using the standard calculation for the relaxation time $t_{relax} = N/(8{\rm ln} N) t_{cross}$.  The mean value for each of these quantities for all sets of parameters is shown in table 1.

The results of these experiments are shown in figures \ref{1000synthesisvirial}--\ref{300synthesiscold}, where we plot the segregation up to the 60 most massive stars for each series of runs.  Note that in these plots the median and its spread are calculated for a single realization, and all ten individual realizations are compared to this one.  While there are small differences in the median distribution between runs, for clarity we show only one, and this does not affect our conclusions.  In table 1, we summarize these plots and show the fraction of runs (out of 10) ${\mathcal N}_i$ that display segregation of the $i$ most massive stars, for $i=5, 10, 20, 40$.  For these purposes we define `displaying segregation' as having a value of $H_c(N)$ above the $+2\sigma$ value for the typical cluster value.  We also plot these fractions for the $n=1000$ cluster in figure \ref{tableexample}.  Examining these figures and the table, it is clear that within a pair of runs with the same age, {\it n}, and $Q_{vir}$ the cluster resulting from a merger displays more segregation, with the difference between solitary Plummer spheres and the merger products being more marked among the higher {\it n} clusters.  As we have chosen to fix the integration time to physical times rather than initial crossing times, this is unsurprising, since the lower-{\it n} clusters are dynamically more mature.  The difference between the Plummer spheres and the merger products is less pronounced with sub-virial initial conditions, and interestingly the $n=1000$ and $n=300$ sub-virial Plummer spheres have identical apparent dynamical ages at the end of the simulations.    

It is also apparent that none of the initial conditions we examine here lead to a cluster displaying the full segregation of the clusters created using the technique of \citet{subr08}, which corresponds to the end-state of dynamical mass segregation.  The segregation in these simulations dominantly affects the most massive stars, a situation that looks more like the ONC data from \citet{hillenbrand97} in figure \ref{ConvexHullExamples}.  These results would then seem to confirm the findings of \citet{mcmillan07} in the lower-{\it n} regime that may be more applicable to local star formation.  As the early evolution of young clusters with imposed primordial segregation across all masses appears to be inconsistent with young observed systems \citep{moeckel09}, this scenario is attractive since it seems to affect only the most massive stars in a cluster over these timescales and population levels.  A final point is that the single Plummer sphere models with cold initial conditions exhibit more segregation of the most massive objects than their virialised counterparts; while not at the same level as the merged clusters, this is another possible route to the formation of dense Trapeziumesque systems.

Examining the pairs of runs of the same physical age in terms of their apparent dynamical age, the general trend is that the merger product appears to be 1--4 dynamical times more advanced than the solitary Plummer sphere.  The exceptions to this are the late-merging sub-virial case with $n=1000$ and the late-merging virial case with $n=300$, both of which appear to be dynamically younger than their solitary counterpart, though the difference is very slight.  The sub-virial initial conditions for the $n=1000$ Plummer spheres actually appear to be dynamically older than their virial, merged counterparts, showing that both merging and sub-virial initial conditions can make a cluster appear older than its true age
  The fraction of runs showing segregation is not directly related to the apparent dynamical age, however; the snapshot afforded by an observation may not reveal the history that led to that point.  In the sub-virial Plummer sphere runs, for instance, the evolution of $\tau_{dyn}$ rises sharply from 0 to $\sim 2.5$ and plateaus for several hundred thousand years, as the cluster collapses, overshoots virial equilibrium, and expands. 
  
  Considering the apparent relaxation times at the end of the simulations, one would not expect complete segregation as a result of two-body relaxation, as none of the systems are older than $t_{relax}$.  Since the segregation timescale for stars of mass $m$ decreases as $<m>/m$, however \citep{spitzer69}, one would expect to see some segregation in the massive stars based on the apparent relaxation times.  The fact that, for instance, the virial Plummer sphere with $n=1000$ and its late-merging counterpart have very similar ages relative to the relaxation time, yet very different mass segregation characteristics, highlights the limited applicability of the relaxation time in young systems of unknown history.
  In more complicated star formation scenarios (i.e. non-spherical, or out of virial equilibrium), observational estimates of the crossing time or relaxation timescale should be treated with caution, as also emphasized by \citet{bastian08} and \citet{allison09b}.

\section{Discussion}
A cautionary note should be sounded about the interpretation of these results and similar early-cluster stellar dynamical experiments.  There have been several attempts in the literature to apply {\it n}-body techniques to very young clusters, when a proper treatment should necessarily include the gas reservoirs out of which the stars are born.  To the extent that the remant gas is treated, it is usually as a spherical or spheroidal background potential on the cluster scale \citep[e.g.][]{adams07,proszkow09}, or as an attempt to interpolate simulated hydrodynamic results on the level of  individual stars \citep{moeckel07b}.  As it becomes more clear from observations and simulations that cluster formation may not be well-described as a spherical process \citep[e.g.][]{myers91,bonnell08,proszkow09}, work from the {\it n}-body end has pushed toward clustered and asymmetrical initial conditions as in \citet{mcmillan07}.  It is not immediately clear that divorcing the stellar dynamics from the gas dynamics and treating effects like mass segregation as purely dynamical processes should be succesful; the inclusion of gas dynamics as in the large-scale simulations as in, for example, \citet{bonnell08} or \citet{bate09} may prove to be obligatory.

The background potential of any remaining gas can be included in an {\it n}-body simulation with relative ease.  Two immediately relevant dissipative effects that cannot be modeled, but are present in hydrodynamic simulations, are the gas drag acting on the stars as they orbit in the cluster potential, and the mass loading that occurs in a competitive accretion scenario.  Both of these effects scale up with the mass of the star, and both should act to accelerate mass segregation.  Interestingly, \citet{bate09} reported minimal segregation\footnote{Though minimal, measurable in the innermost regions; see below.} in his recent simulation that included these effects as well as merging sub-clusters, indicating that (as is usually the case) matters are more complicated than simple prescriptions might allow for.  This suggests that while effects like deceleration due to Bondi-Hoyle accretion could perhaps be added to an {\it n}-body code, the utility of such approximations may be limited.

\begin{figure}
 \includegraphics[width=84mm]{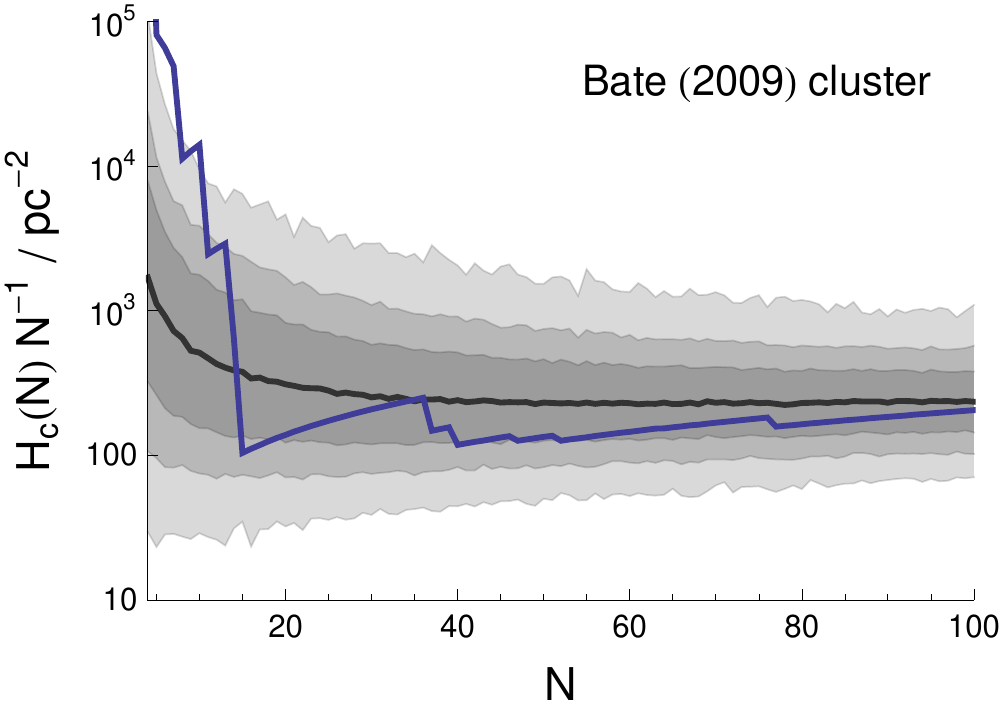}
 \caption{Mass segregation from the end state of the cluster presented in \citet{bate09}. \label{BateCluster}}
\end{figure}

A comparison between pure {\it n}-body merging experiments can be made to the recent hydrodynamic/{\it n}-body simulation presented in \citet{bate09}.  In this simulation, the cluster at $2.85\times10^5$ yr is the end result of the merger of five sub-clusters consisting of $\sim 1250$ stars.  These mergers took place in the presence of the cluster gas.  Bate notes that mass segregation is not present in the end cluster as measured by the cumulative radial distribution, or the median mass, upper quartile mass, and maximum mass except for in the inner 1000 AU.  Applying the segregation detection method used in this paper to the positions of the stars in Bate's simulation, we find evidence of segregation for the most massive stars, shown in figure \ref{BateCluster}.  While there is clearly not general segregation as in the {\v S}ubr clusters, the $\sim 10$ most massive stars display a level of segregation similar in character to the ONC data, and would not stand out as out-of-place in amongst the simulated data in figures \ref{1000synthesisvirial} and \ref{1000synthesiscold}.  While the issue of the equivalence between gas-free and embedded stellar dynamics should be explored further, in this case the level of segregation seen appears to be similar in character.

The mass segregation signal resulting from the merger of sub-clumps of an age and stellar content suitable to nearby young clusters appears to affect predominantly the most massive stars in the cluster, and does not present itself as a general trend that might appear in a cumulative distribution analysis or similar techniques.  Full-scale hydrodynamic simulations \citep{bate09} that feature merging sub-clusters display a similar signal, despite the mergers taking place in the presence of gas.  While the more general type of segregation may have been looked for as the `goal' of segregation-producing phenomena in the past, the type that affects only the most massive stars may be more reflective of Nature (as seen in the ONC), and thus sub-cluster merging and perhaps sub-virial initial conditions seem to be a promising scenario.  

\citet{allison09b} published the results of an independent investigation that overlaps significantly with this work.  Namely, they studied the dynamical effects of sub-virial, clumpy collapse on the mass segregation in a cluster.  Their initial cluster is a more naturalistic setup versus the stylized two-Plummer-sphere clusters we consider here, with 1000 stars placed via a fractal distribution.  The rapid segregation seen in the \citet{allison09b} experiments is mostly due to two-body relaxation in the dense core that develops as the sub-clusters merge and the cluster collapses at early times.  
The extent to which pre-merger segregation in the sub-clusters contributes to the final state (McMillan et al. 2007; Fellhauer et al. 2009; this work, see figure \ref{MergeExample}) versus two-body relaxation in the dense core resulting from clustered, sub-virial collapse \citep{allison09b} is likely dominated by the details of the initial setup.

\begin{figure}
 \includegraphics[width=84mm]{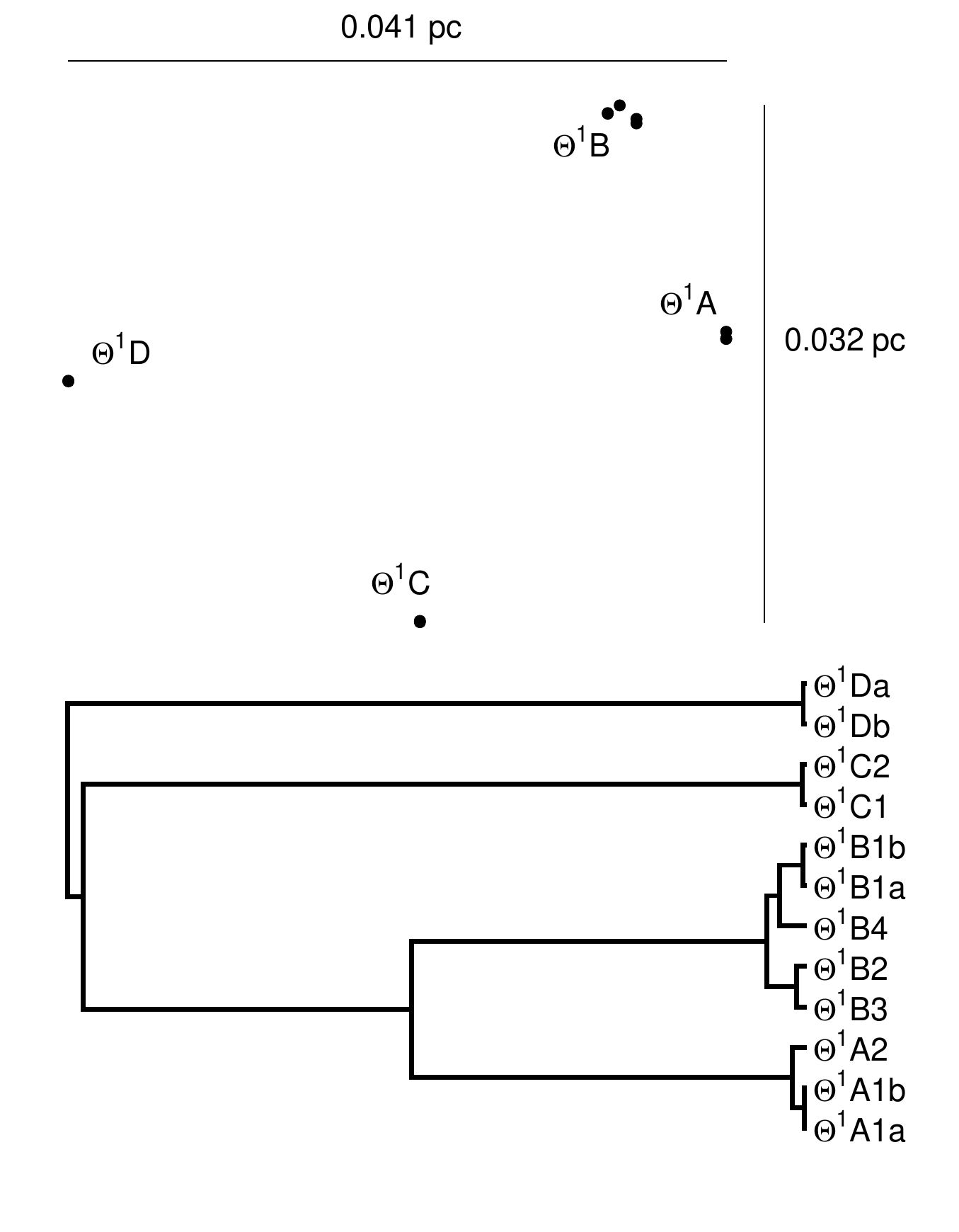}
 \caption{The locations and hierarchy of the Trapezium stars.  The length scales assume a distance of 400 pc. \label{Trapezium}}
\end{figure}

Finally, we caution against an overly-broad definition of Trapezium-like or Trapeziumesque.  Figure \ref{Trapezium} shows the locations and clustering dendrogram of the four OB systems in the Trapezium and their entourage \citep{preibisch99,schertl03,kraus07}.  Most work, including this one, focuses on the four main branches of the dendrogram, i.e. attempting to create a clustering of four or more massive stars.  While this is clearly the first step toward a Trapezium, there are at least two more levels of clustering, in the mini-Trapezium of $\Theta^1{\rm B}$ and the wider binary of the $\Theta^1{\rm A}$ hierarchy, and then the closer visual and spectroscopic binaries of all four main components.  Turning to the younger  W3 IRS 5, with the further complication of surrounding natal gas, there is evidence for multiplicity at sub-1000 AU scales \citep{megeath05}.  Bearing in mind the limited number of examples, multiplicity of the massive sources in Trapezium-like systems seems ubiquitous, and could be fundamental rather than a detail to be worried about later.  If dynamical effects like those considered here and in \citet{allison09b} are to explain the clustering of massive stars, these processes must either leave intact the presumably primordial multiplicity of the massive stars, including higher order subsystems like $\Theta^1{\rm B}$, or produce the multiplicity as part of the segregation; neither possibility has yet been investigated, and consistency with this aspect of the Trapezium may help discriminate between dynamical segregation versus birth location as the cause of massive star clustering.

\section*{Acknowledgments}
Our thanks to Matthew Bate for sharing the stellar positions from his simulation, and the referee for a prompt review.

\begin{table*}
 \centering
 \begin{minipage}{160mm}
  \caption{Cluster characteristics at the end of the integration.  All mean quantities are measured at the end of the simulation.}
  \begin{tabular}{@{}lcccccccccccc@{}}
  \hline
   Cluster type & {\it n} & $ Q_{vir} $& Age & ${\mathcal N}_5$ & ${\mathcal N}_{10}$
     & ${\mathcal N}_{20} $ &${\mathcal N}_{40}$ & $<r_h>$ & $<v_{rms}>$ & $<t_{cross}>$ & $<\tau_{dyn}>$ & Age\\
      & & & (Myr) & & & & & (pc) & (km s$^{-1}$) & (Myr) & &($< t_{relax}>$)\\   
      \hline
 Plummer & 1000  &0.5 & 1.0 &  0.1 & 0.1 &  0.1 & 0.1       &0.40&1.67&0.47&2.1&0.12\\
 Merge late & 1000 & 0.5 & 1.0 &  0.6 & 0.7 & 0.6&0.5       &0.33&1.96&0.33&3.0&0.17\\
 Plummer & 1000 & 0.5 & 0.3  &  0.0 & 0.1 &  0.1 & 0.0        &0.39&1.68&0.45&0.7 &0.04\\
 Merge early&1000 & 0.5 &0.3  &  0.2 & 0.5 &  0.4 & 0.3    &0.27&2.02&0.26&3.8 &0.06\\
 Plummer & 1000 &0.1 & 1.0  &  0.4 & 0.4 &  0.4 & 0.4         &0.27&2.15&0.25&4.0&0.22\\
 Merge late & 1000 & 0.1 & 1.0   &  0.6 & 0.6 &  0.5 & 0.6      &0.31&2.15&0.29&3.5 &0.19\\
 Plummer & 1000 & 0.1 & 0.3 &  0.1 & 0.2 &  0.0 & 0.0         &0.19&2.70&0.14&2.2&0.12\\
 Merge early &1000 & 0.1 & 0.3 &  0.5 & 0.5 &  0.5 & 0.5    &0.10&3.72&0.05&6.0&0.33\\
 Plummer & 300 &0.5 & 1.0 &  0.5 & 0.6 &  0.2 & 0.0             &0.19&1.46&0.25&4.0&0.61\\
 Merge late & 300 & 0.5 & 1.0  &  0.7 & 0.6 &  0.3 & 0.1        &0.21&1.47&0.28&3.5 &0.54\\
 Plummer & 300 & 0.5 & 0.3 &  0.2 & 0.3 &  0.1 & 0.2         &0.16&1.50&0.21&1.5&0.22\\
 Merge early &300 & 0.5 & 0.3 &  0.7 & 0.5 &  0.7 & 0.5        &0.11&1.80&0.13&2.4&0.35 \\
 Plummer & 300 &0.1 & 1.0  &  0.3 & 0.5 &  0.4 & 0.1        &0.25&1.52&0.33&3.0& 0.46\\
 Merge late & 300 & 0.1 & 1.0 & 0.7 & 0.5 &  0.5 & 0.1        &0.22&1.58&0.28&3.5 &0.54\\
 Plummer & 300 & 0.1 & 0.3 &  0.3 & 0.3 &  0.3 & 0.1       &0.10&2.08&0.09&3.3& 0.51\\
 Merge early&300 & 0.1 & 0.3 &  0.7 & 0.4 &  0.4 & 0.4         &0.08&2.40&0.07&4.3&0.65\\
\hline
\end{tabular}
\end{minipage}
\label{summarytable}
\end{table*}

\bibliographystyle{mn2e}	

\begin{thebibliography}{}

\bibitem[\protect\citeauthoryear{{Aarseth}}{{Aarseth}}{2003}]{aarseth03}
{Aarseth} S.~J.,  2003, {Gravitational N-Body Simulations}.
Gravitational N-Body Simulations, by Sverre J.~Aarseth, pp.~430.~ISBN
  0521432723.~Cambridge, UK: Cambridge University Press, November 2003.

\bibitem[\protect\citeauthoryear{{Adams}, {Bloch}, {Butler}, {Druce} \&
  {Ketchum}}{{Adams} et~al.}{2007}]{adams07}
{Adams} F.~C.,  {Bloch} A.~M.,  {Butler} S.~C.,  {Druce} J.~M.,    {Ketchum}
  J.~A.,  2007, \apj, 670, 1027

\bibitem[\protect\citeauthoryear{{Adams}, {Proszkow}, {Fatuzzo} \&
  {Myers}}{{Adams} et~al.}{2006}]{adams06}
{Adams} F.~C.,  {Proszkow} E.~M.,  {Fatuzzo} M.,    {Myers} P.~C.,  2006, \apj,
  641, 504

\bibitem[\protect\citeauthoryear{{Allison}, {Goodwin}, {Parker}, {Portegies
  Zwart}, {de Grijs} \& {Kouwenhoven}}{{Allison} et~al.}{2009a}]{allison09a}
{Allison} R.~J.,  {Goodwin} S.~P.,  {Parker} R.~J.,  {Portegies Zwart} S.~F.,
  {de Grijs} R.,    {Kouwenhoven} M.~B.~N.,  2009a, \mnras, 395, 1449
  
\bibitem[\protect\citeauthoryear{{Allison}, {Goodwin}, {Parker}, {de Grijs},
  {Portegies Zwart} \& {Kouwenhoven}}{{Allison} et~al.}{2009b}]{allison09b}
{Allison} R.~J.,  {Goodwin} S.~P.,  {Parker} R.~J.,  {de Grijs} R.,  {Portegies
  Zwart} S.~F.,    {Kouwenhoven} M.~B.~N.,  2009b, \apjl, 700, L99

\bibitem[\protect\citeauthoryear{{Bastian}, {Gieles}, {Goodwin}, {Trancho},
  {Smith}, {Konstantopoulos} \& {Efremov}}{{Bastian} et~al.}{2008}]{bastian08}
{Bastian} N.,  {Gieles} M.,  {Goodwin} S.~P.,  {Trancho} G.,  {Smith} L.~J.,
  {Konstantopoulos} I.,    {Efremov} Y.,  2008, \mnras, 389, 223

\bibitem[\protect\citeauthoryear{{Bate}}{{Bate}}{2009}]{bate09}
{Bate} M.~R.,  2009, \mnras, 392, 590

\bibitem[\protect\citeauthoryear{{Bonnell}, {Clark} \& {Bate}}{{Bonnell}
  et~al.}{2008}]{bonnell08}
{Bonnell} I.~A.,  {Clark} P.,    {Bate} M.~R.,  2008, \mnras, 389, 1556

\bibitem[\protect\citeauthoryear{{Bonnell} \& {Davies}}{{Bonnell} \&
  {Davies}}{1998}]{bonnell98}
{Bonnell} I.~A.,  {Davies} M.~B.,  1998, \mnras, 295, 691

\bibitem[\protect\citeauthoryear{{Brogan}, {Hunter}, {Indebetouw}, {Chandler},
  {Shirley}, {Rao} \& {Sarma}}{{Brogan} et~al.}{2008}]{brogan08}
{Brogan} C.~L.,  {Hunter} T.~R.,  {Indebetouw} R.,  {Chandler} C.~J.,
  {Shirley} Y.~L.,  {Rao} R.,    {Sarma} A.~P.,  2008, \apss, 313, 53

\bibitem[\protect\citeauthoryear{{Fellhauer}, {Wilkinson} \&
  {Kroupa}}{{Fellhauer} et~al.}{2009}]{fellhauer09}
{Fellhauer} M.,  {Wilkinson} M.~I.,    {Kroupa} P.,  2009, arXiv:0905.0399

\bibitem[\protect\citeauthoryear{{Hillenbrand}}{{Hillenbrand}}{1997}]{hillenbr%
and97}
{Hillenbrand} L.~A.,  1997, \aj, 113, 1733

\bibitem[\protect\citeauthoryear{{Kirk}, {Johnstone} \& {Tafalla}}{{Kirk}
  et~al.}{2007}]{kirk07}
{Kirk} H.,  {Johnstone} D.,    {Tafalla} M.,  2007, \apj, 668, 1042

\bibitem[\protect\citeauthoryear{{Kraus}, {Balega}, {Berger}, {Hofmann},
  {Millan-Gabet}, {Monnier}, {Ohnaka}, {Pedretti}, {Preibisch}, {Schertl},
  {Schloerb}, {Traub} \& {Weigelt}}{{Kraus} et~al.}{2007}]{kraus07}
{Kraus} S.,  {Balega} Y.~Y.,  {Berger} J.-P.,  {Hofmann} K.-H.,  {Millan-Gabet}
  R.,  {Monnier} J.~D.,  {Ohnaka} K.,  {Pedretti} E.,  {Preibisch} T.,
  {Schertl} D.,  {Schloerb} F.~P.,  {Traub} W.~A.,    {Weigelt} G.,  2007,
  \aap, 466, 649

\bibitem[\protect\citeauthoryear{{Lada} \& {Lada}}{{Lada} \&
  {Lada}}{2003}]{lada03}
{Lada} C.~J.,  {Lada} E.~A.,  2003, \araa, 41, 57

\bibitem[\protect\citeauthoryear{{McMillan}, {Vesperini} \& {Portegies
  Zwart}}{{McMillan} et~al.}{2007}]{mcmillan07}
{McMillan} S.~L.~W.,  {Vesperini} E.,    {Portegies Zwart} S.~F.,  2007, \apjl,
  655, L45

\bibitem[\protect\citeauthoryear{{Megeath}, {Wilson} \& {Corbin}}{{Megeath}
  et~al.}{2005}]{megeath05}
{Megeath} S.~T.,  {Wilson} T.~L.,    {Corbin} M.~R.,  2005, \apjl, 622, L141

\bibitem[\protect\citeauthoryear{{Moeckel} \& {Bally}}{{Moeckel} \&
  {Bally}}{2007}]{moeckel07b}
{Moeckel} N.,  {Bally} J.,  2007, \apjl, 661, L183

\bibitem[\protect\citeauthoryear{{Moeckel} \& {Bonnell}}{{Moeckel} \&
  {Bonnell}}{2009}]{moeckel09}
{Moeckel} N.,  {Bonnell} I.~A.,  2009, \mnras, 396, 1864

\bibitem[\protect\citeauthoryear{{Myers}, {Fuller}, {Goodman} \&
  {Benson}}{{Myers} et~al.}{1991}]{myers91}
{Myers} P.~C.,  {Fuller} G.~A.,  {Goodman} A.~A.,    {Benson} P.~J.,  1991,
  \apj, 376, 561

\bibitem[\protect\citeauthoryear{{Peretto}, {Andr{\'e}} \&
  {Belloche}}{{Peretto} et~al.}{2006}]{peretto06}
{Peretto} N.,  {Andr{\'e}} P.,    {Belloche} A.,  2006, \aap, 445, 979

\bibitem[\protect\citeauthoryear{{Pflamm-Altenburg} \&
  {Kroupa}}{{Pflamm-Altenburg} \& {Kroupa}}{2006}]{pflamm-altenburg06}
{Pflamm-Altenburg} J.,  {Kroupa} P.,  2006, \mnras, 373, 295

\bibitem[\protect\citeauthoryear{{Porras}, {Christopher}, {Allen}, {Di
  Francesco}, {Megeath} \& {Myers}}{{Porras} et~al.}{2003}]{porras03}
{Porras} A.,  {Christopher} M.,  {Allen} L.,  {Di Francesco} J.,  {Megeath}
  S.~T.,    {Myers} P.~C.,  2003, \aj, 126, 1916

\bibitem[\protect\citeauthoryear{{Preibisch}, {Balega}, {Hofmann}, {Weigelt} \&
  {Zinnecker}}{{Preibisch} et~al.}{1999}]{preibisch99}
{Preibisch} T.,  {Balega} Y.,  {Hofmann} K.-H.,  {Weigelt} G.,    {Zinnecker}
  H.,  1999, New Astronomy, 4, 531

\bibitem[\protect\citeauthoryear{{Press}, {Teukolsky}, {Vetterling} \&
  {Flannery}}{{Press} et~al.}{2007}]{press07}
{Press} W.~H.,  {Teukolsky} S.~A.,  {Vetterling} W.~T.,    {Flannery} B.~P.,
  2007, {Numerical recipes}.
Cambridge University Press

\bibitem[\protect\citeauthoryear{{Proszkow}, {Adams}, {Hartmann} \&
  {Tobin}}{{Proszkow} et~al.}{2009}]{proszkow09}
{Proszkow} E.-M.,  {Adams} F.~C.,  {Hartmann} L.~W.,    {Tobin} J.~J.,  2009,
  \apj, 697, 1020

\bibitem[\protect\citeauthoryear{{Schertl}, {Balega}, {Preibisch} \&
  {Weigelt}}{{Schertl} et~al.}{2003}]{schertl03}
{Schertl} D.,  {Balega} Y.~Y.,  {Preibisch} T.,    {Weigelt} G.,  2003, \aap,
  402, 267

\bibitem[\protect\citeauthoryear{{Spitzer}}{{Spitzer}}{1969}]{spitzer69}
{Spitzer} L.~J.,  1969, \apjl, 158, L139

\bibitem[\protect\citeauthoryear{{{\v S}ubr}, {Kroupa} \& {Baumgardt}}{{{\v
  S}ubr} et~al.}{2008}]{subr08}
{{\v S}ubr} L.,  {Kroupa} P.,    {Baumgardt} H.,  2008, \mnras, 385, 1673

\bibitem[\protect\citeauthoryear{{Walsh}, {Myers} \& {Burton}}{{Walsh}
  et~al.}{2004}]{walsh04}
{Walsh} A.~J.,  {Myers} P.~C.,    {Burton} M.~G.,  2004, \apj, 614, 194

\bibitem[\protect\citeauthoryear{{Weidner} \& {Kroupa}}{{Weidner} \&
  {Kroupa}}{2006}]{weidner06}
{Weidner} C.,  {Kroupa} P.,  2006, \mnras, 365, 1333

\end{thebibliography}

\bsp

\label{lastpage}

\end{document}